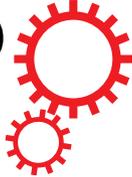

# Effect of lithographically-induced strain relaxation on the magnetic domain configuration in microfabricated epitaxially grown $Fe_{81}Ga_{19}$

R. P. Beardsley[1], D. E. Parkes[1], J. Zemen[2], S. Bowe[1,3], K. W. Edmonds[1], C. Reardon[4], F. Maccherozzi[3], I. Isakov[2,5], P. A. Warburton[5], R. P. Campion[1], B. L. Gallagher[1], S. A. Cavill[3,4] & A. W. Rushforth[1]

We investigate the role of lithographically-induced strain relaxation in a micron-scaled device fabricated from epitaxial thin films of the magnetostrictive alloy $Fe_{81}Ga_{19}$. The strain relaxation due to lithographic patterning induces a magnetic anisotropy that competes with the magnetocrystalline and shape induced anisotropies to play a crucial role in stabilising a flux-closing domain pattern. We use magnetic imaging, micromagnetic calculations and linear elastic modelling to investigate a region close to the edges of an etched structure. This highly-strained edge region has a significant influence on the magnetic domain configuration due to an induced magnetic anisotropy resulting from the inverse magnetostriction effect. We investigate the competition between the strain-induced and shape-induced anisotropy energies, and the resultant stable domain configurations, as the width of the bar is reduced to the nanoscale range. Understanding this behaviour will be important when designing hybrid magneto-electric spintronic devices based on highly magnetostrictive materials.

Many existing and proposed spintronic device concepts make use of magnetic domains and domain walls to store and process data. Examples include magnetoresistive random access memory[1,2], racetrack memory[3,4] and domain wall logic architectures[5]. The drive to develop these technologies has led to a large and growing body of work on the behaviour and structure of magnetic domain configurations and domain walls[6].

The majority of recent studies have used electrical currents to manipulate magnetization, typically by spin transfer torque[7] or by magnetic field[8]. Whilst these methods have received significant attention they have not yet adequately addressed the problem of Joule heating, which presents an increasing problem as logic and memory devices are reduced in size. It has been shown that using electric fields to manipulate magnetization can be many times more efficient than electrical current, due to the achieved reduction in power dissipation within the device[9,10]. Hybrid ferromagnet/piezoelectric devices, in which magnetic anisotropy is controlled by voltage induced strain, are increasingly being seen as a viable and pragmatic solution to the problem of electrical control of magnetization for spintronic applications[11–14]. One of the key elements of such hybrid devices is a magnetostrictive ferromagnetic material, of which $Fe_{81}Ga_{19}$ has the highest magnetostriction coefficient for non-rare-earth containing materials[15], and displays highly magnetostrictive behaviour in the form of thin films[16], making it an excellent candidate for integration into the hybrid structures.

The configuration of ferromagnetic domains and domain walls in a lithographically patterned structure is determined by the balance of the anisotropy energies, including magnetocrystalline, magnetoelastic and shape-induced anisotropy terms. The relative magnitude of these different anisotropies is dependent on device size and aspect ratio. In earlier studies we have shown that the magnetization in large scale (~50 μm) epitaxial $Fe_{81}Ga_{19}$ structures is dominated by magnetoelastic and cubic magnetocrystalline anisotropy terms, and the domains appear disordered with domain walls forming at nucleation sites determined by imperfections in the device structure[12]. In narrower devices, with width ~15 μm and length ~90 μm, shape-induced anisotropy plays a more important role and magnetic domains form a pattern which minimises the stray field from the device[17]. In this letter we discuss an additional contribution to the anisotropy energy which can result from the relaxation of growth strain at the edges of lithographically patterned bars when the width is reduced to the order of 1 μm or less. This strain-relaxation originates from a lattice mismatch between the epitaxially grown magnetic layer and the substrate. The lattice mismatch imposes a built-in compressive strain in the magnetic layer, which can be relaxed by etching of the continuous film into patterned devices. The effect of strain-relaxation on magnetic anisotropy was studied extensively in the diluted magnetic semiconductor (Ga,Mn)As[18–21]. There, the low magnetic moment prevented the formation of regular domain patterns and the observations were interpreted using a single domain model. In our high moment $Fe_{81}Ga_{19}$ devices this additional contribution to the magnetic anisotropy energy results in the stabilisation of a flux closure magnetic domain pattern, which is distorted compared to the pattern observed in wider bars[17] where the effects of the lattice relaxation are less significant.

[1]School of Physics and Astronomy, University of Nottingham, Nottingham NG7 2RD, UK. [2]Department of Physics, Blackett Laboratory, Imperial College, Prince Consort Road, London SW7 2AZ, UK. [3]Diamond Light Source Chilton, Didcot, Oxfordshire OX11 0DE UK. [4]Department of Physics, University of York, Heslington, York, YO10 5DD, UK. [5]London Centre of Nanotechnology, University College London, London, WC1H 0AH, UK. Correspondence and requests for materials should be addressed to S.A.C. (email:stuart.cavill@york.ac.uk) or A.W.R. (email: Andrew.rushforth@nottingham.ac.uk)

The sample consisted of a 14.3 nm $Fe_{81}Ga_{19}$ epitaxial thin film grown by molecular beam epitaxy on a 500 nm Si-doped buffer on a GaAs (001) substrate. A 1.5 nm amorphous GaAs capping layer was grown to protect the metallic layer from oxidation. The layer structure is shown in Fig. 1(a). X-ray diffraction measurements using a Phillips X-Pert materials research diffractometer on material grown under similar conditions[16] show a single peak corresponding to the $Fe_{81}Ga_{19}$ layer, indicating that the layer is a single crystal phase with a vertical lattice parameter of $a_{\perp}^{FeGa} = 0.296$ nm. The lattice constant of the A2 phase of single crystal bulk $Fe_{81}Ga_{19}$ is known to be $a_0^{FeGa} = 0.287$ nm[22]. Taking the in plane lattice constant of a fully strained film to be half the GaAs substrate lattice parameter ($a_{\parallel}^{FeGa} = 0.5 a_0^{GaAs} = 0.283$ nm) and assuming a Poisson ratio of 0.45[23] this would imply that the out of plane lattice constant of a fully strained $Fe_{81}Ga_{19}$ film on GaAs would be 0.296 nm, consistent with the measured value. Superconducting quantum interference device (SQUID) magnetometry shows that the cubic magnetocrystalline anisotropy favouring the [100]/[010] directions has a magnitude $K_C = 18.9 \times 10^3$ J/m$^3$ for the unpatterned film, and the weaker uniaxial anisotropy favouring the [110] direction is $K_U = 12.4 \times 10^3$ J/m$^3$. The latter contribution is typically observed for ferromagnetic films grown on GaAs substrates and is induced by the substrate/film interface[24,25]. The saturation magnetisation of the film is $1.4 \times 10^6$ A/m. An L-shaped structure with arms of width 1.2 μm and length 10 μm along the [100]/[010] directions was fabricated using electron beam lithography and Ar ion milling. The milled depth was greater than the film thickness which resulted in a 100 nm GaAs mesa, measured by atomic force microscopy, with the $Fe_{81}Ga_{19}$ and capping layers on top (Fig. 1(a)). In this letter we focus on the behaviour in one 10 μm-long arm of the L-shaped structure.

Finite element calculations were used to gain insight into the structural behaviour of a bar of infinite length and the same cross-sectional dimensions as the experimental device. A fixed structural constraint was set at the substrate boundaries and an initial in-plane compressive strain of $(a_0^{FeGa} - a_{\parallel}^{FeGa})/a_0^{FeGa} = 1.4\%$ was included in the $Fe_{81}Ga_{19}$ layer. The strain profiles in the wire cross section were calculated using a partial differential equation solver (the COMSOL software package) implementing the theory of an anisotropic elastic medium. From here on we define positive strain as the difference between the in-plane lattice spacing along the directions perpendicular and parallel to the stripe such that the zero strain state corresponds to the unrelaxed film.

Figure 1(a,b) show the calculated strain relaxation across the wire cross-section, defined as $\varepsilon_{xx} - \varepsilon_{yy}$ where $\varepsilon_{xx}$ and $\varepsilon_{yy}$ are the in-plane components of strain in directions perpendicular and parallel to the wire. The strain profile at the $Fe_{81}Ga_{19}$/cap interface, shown in Fig. 1(c), reveals that there is a nonzero strain relaxation in the centre of the cross-section, which increases in amplitude away from the bar centre. The relaxation in the edge-region was also seen in the previous studies on (Ga,Mn)As-based devices[18–21]. An interesting feature of this profile is the abrupt decrease in amplitude in the regions near to the edges of the bar. This discontinuity is only observed with the inclusion of the GaAs capping layer, which tends to suppress the relaxation of the in-built strain by partially clamping the top surface as shown in Fig. 1(b).

Magnetic domains were imaged using photoemission electron microscopy (PEEM) on beamline I06 of the Diamond Light Source[26]. Illuminating the sample at oblique incidence and making use of X-ray magnetic circular dichroism at the Fe $L_3$ edge as the contrast mechanism allowed sensitivity to in-plane moments with a spatial resolution of approximately 50 nm. Figure 2(a) shows an image of the domain configuration in a 1.2 μm × 6 μm section of the bar. The flux-closure domain configuration observed is different to that seen in previous studies of wires with width 15 μm[17] in that the regions with magnetisation perpendicular to the length of the bar are broadened at the edges of the bar. In this study there is no externally induced strain and we attribute the observed domain behaviour to the relatively large effect of non-linear strain relaxation at the bar edges in our narrower device.

To understand the experimentally observed domain configuration we performed micromagnetic calculations carried out using the Object Oriented Micromagnetic Framework (OOMMF)[27]. The simulation used magnetocrystalline anisotropy coefficients determined by the SQUID magnetometry measurements of the unpatterned $Fe_{81}Ga_{19}$ film, a cell size of 1 nm × 1 nm × 10 nm and critical damping. The OOMMF simulation was initialised in a flux-closing state, with flux-closing units having an aspect ratio of 1:1.3, which is the average aspect ratio present in the experimental image.

The magneto-elastic coupling present in the $Fe_{81}Ga_{19}$ film leads to a strain-induced uniaxial anisotropy across the width of the bar, with a profile determined by the strain profile shown in Fig. 1(c). The relation between strain and magneto-elastic anisotropy energy is, $\Delta K_{me} = B_{me}\varepsilon$, where $\Delta K_{me}$ represents the change in the magnetic anisotropy energy and $\varepsilon$ is the position-dependent uniaxial strain perpendicular to the bar length. We set the magneto-elastic constant, $B_{me} = 1.56 \times 10^7$ J/m$^3$, as determined previously for an epitaxial thin $Fe_{81}Ga_{19}$ film[16]. We have approximated the anisotropy energy as a function of position by fitting a first order polynomial to the edge region, and an exponential function to the central region of the calculated strain profile. This strain-induced anisotropy energy was incorporated into the OOMMF calculation as an additional uniaxial magnetocrystalline anisotropy term, with the anisotropy axis perpendicular to the length of the bar. The results of micromagnetic calculations based on the approximated anisotropy profile are shown in Fig. 2(b). Similar to the experimental data, the ground state is a flux-closure pattern with regions magnetised perpendicular to the length of the bar. To observe broadening of the domain boundaries at the edges of the bar to an extent similar to that observed in the experimental data, we scale the magnitude of the strain-induced anisotropy by a factor of 0.4 in the simulations, otherwise the calculated broadening is too large. A possible explanation for the need to scale the anisotropy might arise from damage to the $Fe_{81}Ga_{19}$ layer at the edges of the bar during device fabrication, which would degrade the magnetism in the region where the strain relaxation is largest. If the strain-induced anisotropy is not included in the calculation we find that the flux closure domain configuration is not the lowest energy state of the system. Calculations initialised with a single domain state, where magnetisation is aligned uniformly along the length of the bar, evolve into the S-shaped domain pattern shown in Fig. 2(c). In the absence of the strain-induced anisotropy the S-shaped pattern represents a lower total energy state than the flux closure state. The situation is reversed when the strain-induced anisotropy is included in the calculation. Calculations on bars with the same 1.2 μm width, but using periodic boundary conditions to simulate infinite length reveal that a very similar flux closure pattern represents the lowest energy ground state when the strain-induced anisotropy is included, but that without this anisotropy term a single domain configuration with magnetisation pointing along the length of the bar is the lowest energy configuration.

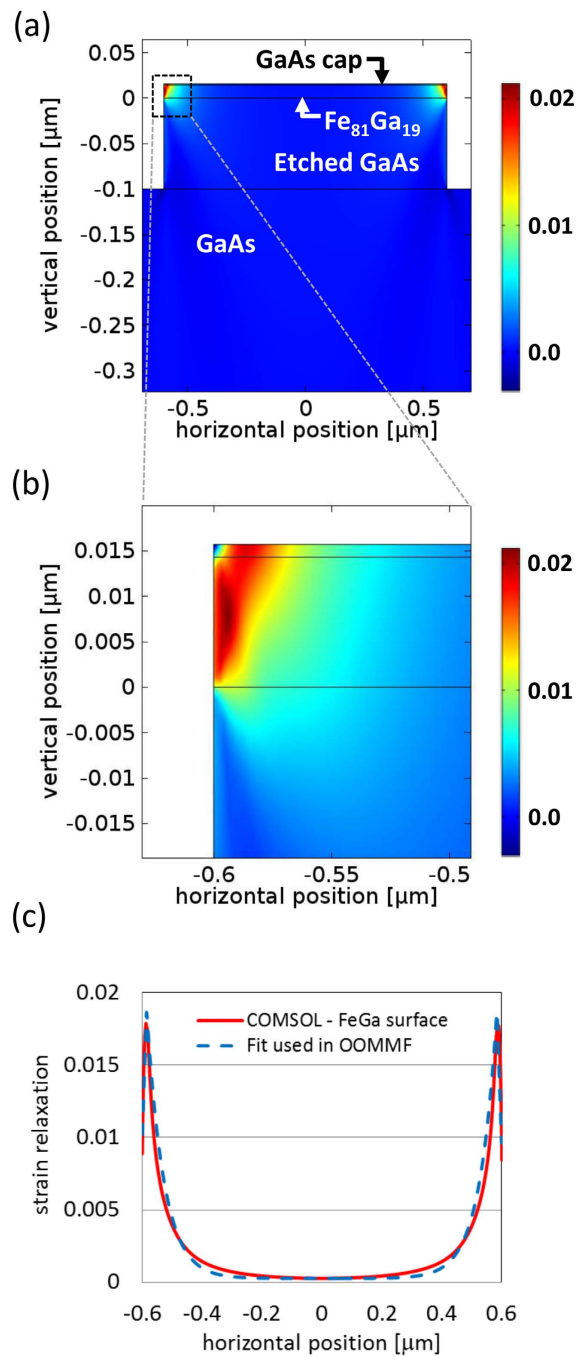

**Figure 1. The calculated strain relaxation across the micro-bar.** (**a**) Cross-sectional view of the layer structure of the experimental device and simulated colour scale map showing the relaxation of the growth strain as a function of depth in the bar. (**b**) A zoomed section of the colour scale map showing the relaxation of the growth strain as a function of depth in the edge region of the bar. (**c**) The simulated strain profile across the $Fe_{81}Ga_{19}$ bar at the $Fe_{81}Ga_{19}$/cap interface.

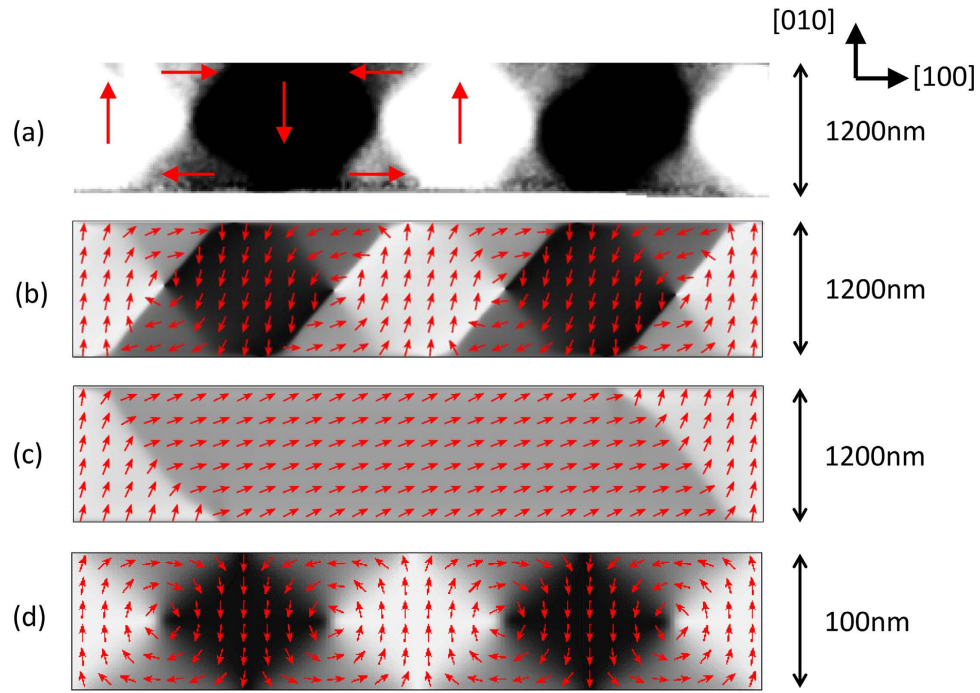

**Figure 2. The magnetic domain configuration.** (**a**) Experimental top down PEEM image of 1.2 μm × 6 μm region of a bar with arrows indicating the magnetization direction. (**b**) Micromagnetic simulation for the 1.2 μm × 6 μm bar with an anisotropy profile that includes the calculated strain relaxation profile scaled by a factor of 0.4. (**c**) Micromagnetic simulation for the 1.2 μm × 6 μm bar initialised in a single domain configuration without the inclusion of a strain-induced anisotropy. (**d**) As in (**b**), but for a 100 nm × 500 nm bar.

To investigate the competition between the shape- and strain-induced anisotropies and to determine the limit in which shape-induced anisotropy will overcome the strain-induced anisotropy, we carried out calculations for bars of different widths, but with the same length to width ratio, thickness and etch depth as in the calculations described above. Figure 3(a) shows the calculated strain profile as a function of the normalised position across the bar. It can be observed that the maximum strain at the edges of the bar, where the relaxation is largest, remains roughly the same for each bar width. The strain relaxation at the centre of the bar, and therefore also the average strain relaxation across the bar, becomes larger as the bar width decreases. Figure 3(b) shows the difference, $\Delta E$, between the total energies of the flux closure and S-shaped states as a function of bar width. For widths in the range 100 nm to 2000 nm (300 nm or greater and less than 2000 nm for the infinitely long bar), the flux closure state is the energetically favourable state when the strain-induced anisotropy is included in the calculation. For widths greater than 500 nm (300 nm for the infinitely long bar) the magnitude of $\Delta E$ decreases as width increases, representing the reducing significance of the strain-induced anisotropy which acts mainly at the edges of the bar. Below 500 nm (300 nm for the infinitely long bar) the magnitude of $\Delta E$ decreases as the bar width decreases until eventually the S-shaped state becomes energetically more favourable. This transition occurs below a width of 100 nm (300 nm for the infinitely long bar) and represents the increasing significance of the demagnetising field with respect to the strain-induced anisotropy as the width of the bar is reduced to these dimensions. For the 100 nm wide bar the flux closure domain pattern is the energetically favoured state. In this case the strain relaxation is significant across the whole width of the bar. However, Fig. 2(d) reveals that the broadening of the transverse domains at the edges of the bar is reduced compared to the case of the 1200 nm bar (Fig. 2(b)) due to the increased significance of the demagnetising field which competes with the strain-induced anisotropy. We note that, although not considered in the present study, it would be important to include the effects of the strain-induced anisotropy in the length direction of the bar for devices with dimensions in the sub-micron limit. In the absence of strain-induced anisotropy the S-shaped state is energetically favourable for both the finite and infinite length bars over the whole range of widths considered.

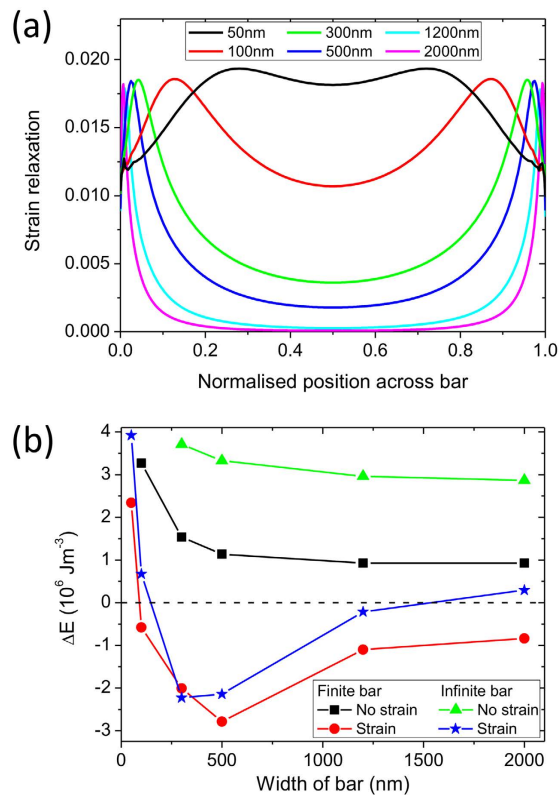

**Figure 3. The dependence on the width of the bar.** (**a**) Calculated strain profile across the $Fe_{81}Ga_{19}$ bar at the $Fe_{81}Ga_{19}$/cap interface as a function of the normalised position across the bar for different widths. (**b**) The difference in the total energy calculated for a bar initialised in the flux closure domain state and a bar with the S-shaped domain state, $\Delta E$, as a function of the width of the bar. The dashed line represents $\Delta E = 0$.

In our $Fe_{81}Ga_{19}$ structures the cubic magnetocrystalline anisotropy supports magnetic easy axes along the [100] and [010] directions. The small intrinsic uniaxial anisotropy along the [110] direction acts to distort the shape of the magnetic domains and leads to a canting of the magnetic moments towards the [110] direction. This feature is present in the experimental data (Fig. 2(a)) and is also revealed in the calculations. Furthermore, the magnetostriction constant in $Fe_{81}Ga_{19}$ is largest along the [100]/[010] directions, hence magneto-elastic effects will be maximised by strain relaxation along these directions[23], as is the case with our device.

Magnetic contrast imaging by Kerr microscopy on patterned and unpatterned films reveals no evidence of the formation of magneto-statically and magneto-elastically self-sufficient domains, which were reported by Chopra *et al.* for $Fe_xGa_{1-x}$ single crystals after high-temperature thermal processing[28]. This may be due to the different growth method used in our study or the fact that our epitaxial films are clamped to a thick substrate. In our case, a micromagnetic model incorporating strain relaxation induced anisotropy energy is sufficient to understand the experimental observations.

In conclusion we have demonstrated that relaxation of growth strain is an important factor in determining the magnetic domain configuration of micron and sub-micron sized devices based on epitaxial $Fe_{81}Ga_{19}$. In the 1.2 μm wide bars investigated experimentally, the strain-induced anisotropy stabilises the formation of a regular flux-closure domain configuration and distorts the features near the edges of the bar. The competition between strain- and shape-induced anisotropy energies determines the stable domain configuration over a range of device dimensions. Growth strain is an additional degree of freedom to be considered and manipulated in the design of micro- and nano-scale magnetic devices.

## References

1. Fukami, S. *et al.* Low-current perpendicular domain wall motion cell for scalable high-speed MRAM. 2009 Symposium on *VLSI Technology Digest of Technical Papers* (2009).
2. Gallagher, W. J. & Parkin, S. S. P. Development of the magnetic tunnel junction MRAM at IBM: From first junctions to a 16-Mb MRAM demonstrator chip. *IBM J. Res. Dev*. **50,** 5–23 (2006).
3. Parkin, S. S. P. Racetrack memory device. US Patent 6834005 (2012).
4. Parkin, S. S. P., Hayashi, M. & Thomas, L. Magnetic Domain-Wall Racetrack Memory. *Science* **320,** 190–194 (2008).
5. Allwood, D. A., Xiong, G., Faulkner, C. C., Atkinson, D., Petit, D. & Cowburn, R. P. Magnetic Domain-Wall Logic. *Science* **309,** 1688–1692 (2005).
6. Ruediger, U., Yu, J., Zhang, S., Kent, A. D. & Parkin, S. S. P. Negative Domain Wall Contribution to the Resistivity of Microfabricated Fe Wires. *Phys. Rev. Lett*. **80,** 5639–5642 (1998).
7. De Ranieri, E. *et al.* Piezoelectric control of the mobility of a domain wall driven by adiabatic and non-adiabatic torques. *Nat. Mater*. **12,** 808–814 (2013).

8. Beach, G. S. D., Nistor, C., Knutson, C., Tsoi, M. & Erskine, J. L. Dynamics of field-driven domain-wall propagation in ferromagnetic nanowires. *Nat. Mater.* **4,** 741–744 (2005).
9. Hu, J.-M., Li, Z., Chen, L.-Q. & Nan, C.-W. High-density magnetoresistive random access memory operating at ultralow voltage at room temperature. *Nat. Commun.* **2,** 553 (2011).
10. Roy, K., Bandyopadhyay, S. & Atulasimha, J. Hybrid spintronics and straintronics: A magnetic technology for ultra low energy computing and signal processing. *Appl. Phys. Lett.* **99,** 063108 (2011).
11. Lei, N. *et al.* Strain-controlled magnetic domain wall propagation in hybrid piezoelectric/ferromagnetic structures. *Nat. Commun.* **4,** 1378 (2013).
12. Parkes, D. E. *et al.* Non-volatile voltage control of magnetization and magnetic domain walls in magnetostrictive epitaxial thin films. *Appl. Phys. Lett.* **101,** 072402 (2012).
13. Ahmad, H., Atulasimha, J. & Bandyopadhyay, S. Reversible strain-induced magnetization switching in FeGa nanomagnets: Pathway to a rewritable, non-volatile, nontoggle, extremely low energy straintronic memory. *Sci. Rep.* **5,** 18264 (2015).
14. Ahmad, H., Atulasimha, J. & Bandyopadhyay, S. Electric field control of magnetic states in isolated and dipole-coupled FeGa nanomagnets delineated on a PMN-PT substrate. *Nanotechnology* **26,** 401001 (2015).
15. Clark, A. E., Wun-Fogle, M., Restorff, J. B. & Lograsso, T. A. Magnetostrictive properties of Galfenol alloys under compressive stress. *Materials transactions* **43,** 881–886 (2002).
16. Parkes, D. E. *et al.* Magnetostrictive thin films for microwave spintronics. *Sci. Rep.* **3,** 2220, (2013).
17. Cavill, S. A. *et al.* Electrical control of magnetic reversal processes in magnetostrictive structures. *Appl. Phys. Lett.* **102,** 032405 (2013).
18. Wunderlich, J. *et al.* Local control of magnetocrystalline anisotropy in (Ga,Mn)As microdevices: Demonstration in current-induced switching. *Phys. Rev. B* **76,** 054424 (2007).
19. Wenisch, J. *et al.* Control of Magnetic Anisotropy in (Ga,Mn)As by Lithography-Induced Strain Relaxation. *Phys. Rev. Lett.* **99,** 077201 (2007).
20. King, C. S. *et al.* Strain control of magnetic anisotropy in (Ga,Mn)As microbars. *Phys. Rev. B* **83,** 115312 (2011).
21. Hoffmann, F., Woltersdorf, G., Wegscheider, W., Einwanger, A., Weiss, D. & Back, C. H. Mapping the magnetic anisotropy in (Ga,Mn)As nanostructures. *Phys. Rev. B* **80,** 054417 (2009).
22. Bhattacharyya, S., Jinschek, J. R., Khachaturyan, A., Cao, H., Li, J. F. & Viehland, D. Nanodispersed DO3-phase nanostructures observed in magnetostrictive Fe-19% Ga Galfenol alloys. *Phys. Rev. B* **77,** 104107 (2008).
23. Clark, A. E. *et al.* Extraordinary magnetoelasticity and lattice softening in bcc Fe-Ga alloys. *J. Appl. Phys.* **93,** 8621–8623 (2003).
24. Wastlbauer, G. & Bland, J. A. C. Structural and magnetic properties of ultrathin epitaxial Fe films on GaAs(001) and related semiconductor substrates. *Advances in Physics* **54,** 137 (2005).
25. Hindmarch, A. T., Rushforth, A. W., Campion, R. P., Marrows, C. H. & Gallagher, B. L. Origin of in-plane uniaxial magnetic anisotropy in CoFeB amorphous ferromagnetic thin films. *Phys. Rev. B* **83,** 212404 (2011).
26. Dhesi, S. S. *et al.* The Nanoscience Beamline (I06) at Diamond Light Source. *AIP Conference Proceedings* **1234,** 311–314 (2010).
27. Donahue, M. J. & Porter, D. G. OOMMF User's Guide, Version 1, Interagency Report NISTIR 6376 (1999).
28. Chopra, H. D. & Wuttig, M. Non-Joulian magnetostriction. *Nature*, **521,** 340 (2015).

## Acknowledgements

The authors would like to acknowledge Diamond Light Source for the provision of beamtime under SI-8560 and 7601. This work was supported by the Engineering and Physical Sciences Research Council [grant number EP/H003487/1]. We are grateful for access to the University of Nottingham High Performance Computing Facility.

## Author Contributions

R.B., D.E.P., S.B., K.W.E., F.M., S.A.C. and A.W.R. conducted the PEEM measurements. R.B., D.E.P. and C.R. performed electron beam lithography. II carried out ion milling. J.Z. performed structural calculations of the strain profile using the COMSOL package. D.E.P. and A.W.R. carried out the micromagnetic simulations. R.P.C. was responsible for the growth of the semiconductor and metallic layers. All authors contributed to writing the manuscript.

## Additional Information

**Competing financial interests:** The authors declare no competing financial interests.

**How to cite this article**: Beardsley, R. P. *et al.* Effect of lithographically-induced strain relaxation on the magnetic domain configuration in microfabricated epitaxially grown $Fe_{81}Ga_{19}$. *Sci. Rep.* **7**, 42107; doi: 10.1038/srep42107 (2017).

**Publisher's note:** Springer Nature remains neutral with regard to jurisdictional claims in published maps and institutional affiliations.